\documentclass[journal]{IEEEtran}
\usepackage{graphicx}
\usepackage{color}
\usepackage{refstyle}
\usepackage{amsthm,amssymb}
\usepackage{amsmath}
\usepackage{algorithm}
\usepackage{algorithmic}
\usepackage{epstopdf}
\usepackage{mathtools,amssymb}
\usepackage{cuted}

\begin{document}
\linespread{1}
\title{Study of Tomlinson-Harashima Precoding Strategies for Physical-Layer Security in Wireless Networks }

\author{Xiaotao Lu, Rodrigo C. de~Lamare and Keke Zu
\thanks{Xiaotao Lu is with the Communications Research Group, Department of Electronics,
University of York, YO10 5DD York, U.K., R. C. de Lamare is with CETUC, PUC-Rio, Brazil and with the
Communications Research Group, Department of Electronics, University
of York, YO10 5DD York, U.K. and Keke Zu is with Ericsson Research,
Sweden, e-mails: xtl503@york.ac.uk;
rodrigo.delamare@york.ac.uk;zukeke@gmail.com.}}

\maketitle
\begin{abstract}
In this paper, we propose novel non-linear precoders for the
downlink of a multi-user MIMO system with the existence of multiple
eavesdroppers. The proposed non-linear precoders are designed to
improve the physical-layer secrecy rate. Specifically, we combine
the non-linear successive optimization Tomlinson-Harashima precoding
(SO-THP) with generalized matrix inversion (GMI) technique to
maximize the physical-layer secrecy rate. For the purpose of
comparison, we examine different traditional precoders with the
proposed algorithm in terms of secrecy rate as well as BER
performance. We also investigate simplified generalized matrix
inversion (S-GMI) and lattice-reduction (LR) techniques in order to
efficiently compute the parameters of the precoders. We further
conduct computational complexity and secrecy rate analysis of the
proposed and existing algorithms. In addition, in the scenario
without knowledge of channel state information (CSI) to the
eavesdroppers, a strategy of injecting artificial noise (AN) prior
to the transmission is employed to enhance the physical-layer
secrecy rate. Simulation results show that the proposed non-linear
precoders outperform existing precoders in terms of BER and secrecy
rate performance.
\end{abstract}

\section{Introduction}

Data security in wireless systems has been traditionally dominated
by encryption methods such as Data Encryption Standard (DES) and
Advanced Encryption Standard (AES) \cite{cyber}. However, these
{existing} encryption algorithms suffer from high complexity and
{high} latency. {Besides, development in computing power also brings
great challenges to existing encryption techniques.} Therefore, are
capable of achieving secure transmission under high computing power
scenario with low complexity have become an important research
topic.

From the viewpoint of information theory, Shannon established the
theorem of cryptography in his seminal paper \cite{Shannon}. Wyner
has subsequently posed the Alice-Bob-Eve problem and described the
wire-tap transmission system \cite{wyner}. Furthermore, the system
discussed in \cite{wyner} suggests that physical layer security can
be achieved in wireless networks. Later on, another study reported
in \cite{csiszar} proved that secrecy transmission is achievable
even under the situation that the eavesdropper has a better channel
than the desired user in a statistical sense. Furthermore, the
secrecy capacity for different kinds of channels, such as Gaussian
wire-tap channel and multi-input multi-output (MIMO) wire-tap
channel have been studied in \cite{hellman,oggier}. In some later
works \cite{goel,mukherjee,lu2015opportunistic}, it has been found
that the secrecy of the transmission can be further enhanced by
adding artificial noise to the system.

\subsection{Prior and Related Work}

In recent years, precoding techniques, which rely on knowledge of
channel state information (CSI), have been widely studied in the
downlink of multiuser MIMO (MU-MIMO) and cooperative systems
\cite{mmimo,wence,Costa,delamare_ieeproc,TDS_clarke,TDS_2,peng2016adaptive,switch_int,switch_mc,smce,TongW,jpais_iet,TARMO,keke1,kekecl,keke2,Tomlinson,dopeg_cl,peg_bf_iswcs,gqcpeg,peg_bf_cl,healy2016design,Harashima,mbthpc,zuthp,rmbthp,cai2015robust,Hochwald,BDVP},
\cite{delamare_mber,rontogiannis,delamare_itic,stspadf,choi,stbcccm,FL11,jio_mimo,peng_twc,spa,spa2,jio_mimo,P.Li,jingjing,did,bfidd,mbdf,gu2016joint,cai2015adaptive,arevalo2016variable}.
. Linear precoding techniques such as zero-forcing (ZF), minimum
mean-square error (MMSE) and block diagonalization (BD) have been
introduced and studied in \cite{stankovic,cai,geraci}. Furthermore,
non-linear precoding techniques like Tomlinson-Harashima precoding
(THP) \cite{payaro}, vector perturbation (VP) precoding
\cite{sebdani} have also been reported and investigated. In the
previous mentioned works, the implementation of linear or non-linear
precoding techniques at the transmitter are considered with perfect
knowledge of CSI to the users. In the scenario without knowledge of
CSI to the eavesdroppers, one technique which is effective in
improving the secrecy rate of the downlink of MU-MIMO systems is the
application of artificial noise (AN) at the transmitter \cite{goel}.
Several criteria or strategies applying AN to wireless systems have
been introduced in \cite{lin,chorti}. In particular, the approaches
reported in \cite{mukherjee} have been applied to the downlink of
MU-MIMO systems. Apart from the studies in precoding techniques
there are also some works that introduce lattice-reduction (LR)
strategies \cite{yao,taherzadeh}. The LR strategies are also
implemented prior to the transmission and it has been proved that
the LR aided system can achieve full diversity in the downlink of
MU-MIMO systems.

\subsection{Motivation and Contributions}

Prior work on precoding for physical-layer security systems has been
heavily based on \cite{goel,mukherjee}, which can effectively
improve the secrecy rate of wireless systems. However, it is well
known in the wireless communications literature that non-linear
precoding techniques can outperform linear approaches. In
particular, non-linear precoding techniques require lower transmit
power than linear schemes and can achieve higher sum-rates. However,
work on non-linear precoding for physical-layer security in wireless
systems is extremely limited even though there is potential to
significantly improve the secrecy rate of wireless systems. The
motivation for this work is to develop and study non-linear
precoding algorithms for MU-MIMO systems that can achieve a secrecy
rate higher than that obtained by linear precoders as well as a
lower transmit power requirement and an improved bit error rate
(BER) performance.

In our work, we develop and study a successive optimization
Tomlinson-Harashima precoding (SO-THP) precoding algorithms based on
the generalized matrix inversion approach reported in \cite{geraci}.
Specifically, the proposed non-linear precoders exploit both
successive interference cancellation, lattice-reduction and block
diagonalization, which can impose orthogonality between the channels
of the desired users
\cite{scharf,bar-ness,pados99,reed98,hua,goldstein,santos,qian,delamarespl07,xutsa,delamaretsp,kwak,xu&liu,delamareccm,wcccm,delamareelb,jidf,delamarecl,delamaresp,delamaretvt,jioel,delamarespl07,delamare_ccmmswf,jidf_echo,delamaretvt10,delamaretvt2011ST,delamare10,fa10,lei09,ccmavf,lei10,jio_ccm,ccmavf,stap_jio,zhaocheng,zhaocheng2,arh_eusipco,arh_taes,dfjio,rdrab,dcg_conf,dcg,dce,drr_conf,dta_conf1,dta_conf2,dta_ls,song,wljio,barc,jiomber,saalt}.
. This combined approach has not been considered previously in the
literature and has the potential of achieving a higher secrecy rate
than existing non-linear and linear precoding algorithms as well as
an improved BER performance as compared to prior art. The major
contributions in our paper are summarized as follows:
\begin{itemize}

  \item A novel non-linear precoding technique, namely, SO-THP+GMI
  is proposed for the downlink of MU-MIMO networks in the presence of multiple eavesdroppers.

\item The proposed
  SO-THP+GMI algorithm combines the SO-THP precoding with
  the GMI technique to achieve a higher secrecy rate.

  \item The proposed SO-THP+GMI precoding algorithm is extended
  to a S-GMI version which aims to reduce computational complexity of the SO-THP+GMI algorithm.

  \item An LR strategy is combined with the aforementioned S-GMI
  version proposed algorithm and this so-called LR-aided version
  algorithm achieves full receive diversity.

  \item An analysis of the secrecy rate achieved by the proposed
  non-linear precoding algorithms is carried out along with an
  assessment of their computational complexity cost.

\end{itemize}

The rest of this paper is organized as follows. We begin in
Section~\ref{sec:System} by introducing the system model and the
performance metrics. A brief review of the standard SO-THP algorithm
is included in Section~\ref{sec:Review}. In
Section~\ref{sec:Algorithms}, we present the details of proposed
SO-THP+GMI, SO-THP+S-GMI and LR-SO-THP+S-GMI precoding algorithms.
Next in Section~\ref{sec:Analysis}, the analysis of secrecy rate and
the computational complexity of the precoding algorithms are carried
out. In Section~\ref{sec:Simulation}, numerical evaluation is
conducted to show the advantage of proposed precoding algorithms.
Finally, some concluding remarks are given in
Section~\ref{sec:Conclusion}.

\subsection{Notation}
Bold uppercase letters ${\boldsymbol A}\in {\mathbb{C}}^{M\times N}$
denote matrices with size ${M\times N}$ and bold lowercase letters
${\boldsymbol a}\in {\mathbb{C}}^{M\times 1}$ denote column vectors
with length $M$. Conjugate, transpose, and conjugate transpose are
represented by $(\cdot)^\ast$, $(\cdot)^T$ and $(\cdot)^H$
respectively; $\boldsymbol I_{M}$ is the identity matrix of size
$M\times M$; $\rm diag \{\boldsymbol a\}$ denotes a diagonal matrix
with the elements of the vector $\boldsymbol a$ along its diagonal;
$\mathcal{CN}(0,\sigma_{n}^{2})$ represents complex Gaussian random
variables with $i.i.d$ entries with zero mean and $\sigma_{n}^{2}$
variance.

\section{System Model and Performance Metrics}
\label{sec:System}

In this section we introduce the system model of the downlink of the
MU-MIMO network under consideration. The performance metrics used in
the assessment of the proposed and existing techniques are also
described.

\subsection{System Model}

\begin{figure}[h]
\centering
\includegraphics[scale=0.6]{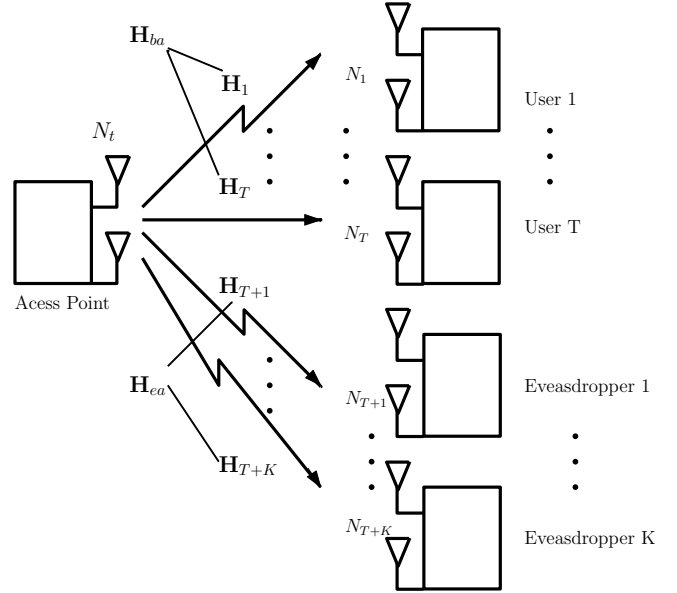}
\caption{System model of a MU-MIMO system with T users and K eavesdroppers}
\label{fig:sys}
\end{figure}

Consider a MU-MIMO downlink wireless network consisting of one
transmitter or Alice at the access point, T users or Bob and K
eavesdroppers or Eve at the receiver side as shown in
Fig.~\ref{fig:sys}. The transmitter is equipped with $N_{t}$
antennas. Each user and each eavesdropper node are equipped with
$N_{r}$ and $N_{k}$ receive antennas, respectively. In this system
we assume that the eavesdroppers do not jam the transmission and the
channel from the transmitter to each user or eavesdropper follows a
flat-fading channel model. The quantities ${\boldsymbol H}_{r}\in
{\mathbb{C}}^{N_{r}\times N_{t}}$ and ${\boldsymbol H}_{k}\in
{\mathbb{C}}^{N_{k}\times N_{t}}$ denote the channel matrix of the
$i$th user and $k$th eavesdropper, respectively. Following
\cite{haardt}, the number of antennas should satisfy $N_{t}^{\rm
total} \geqslant T \times N_{r}$. During the transmission, $N_{t} =
M \times N_{r}$ antennas at the transmitter are activated to perform
the precoding procedure. In other words, the precoding matrix is
assumed here for convenience to be always a square matrix.

We use the vector ${\boldsymbol s}_{r}\in {\mathbb{C}}^{N_{r}\times
1}$ to represent the data symbols to be transmitted to user $r$. An
artificial noise (AN) can be injected before the data transmission
to enhance the physical-layer secrecy. We use the vector
${\boldsymbol s}_{r}'\in {\mathbb{C}}^{m\times 1}$ with $m \leqslant
(N_{t}^{\rm total}-T \times N_{r})$ to denote the independently
generated jamming signal. Assume the transmit power of user $r$ is
$E_{r}$, and $0<\rho<1$ is the power fraction devoted to the user.
Then, the power of user and jamming signal can be respectively
expressed as $E[\boldsymbol s_{r}^{H}\boldsymbol s_{r}]=\rho E_{r}$
and $E[\boldsymbol s_{r}'^{H}\boldsymbol s_{r}']=(1-\rho) E_{r}$.
Finally the signal after precoding can be expressed as
\begin{equation}
{\boldsymbol x}_{r}={\boldsymbol P}_{r}{\boldsymbol s}_{r}+
{\boldsymbol P}_{r}' {\boldsymbol s}_{r}',
\label{eqn:xr2}
\end{equation}
where the quantities ${\boldsymbol P}_{r}\in
{\mathbb{C}}^{N_{t}\times N_{r}}$ and and ${\boldsymbol P}_{r}'\in
{\mathbb{C}}^{N_{t}\times m}$ are the corresponding precoding
matrices. Here we take zero-forcing precoding as an instance. Given
the total channel matrix $\boldsymbol H=[\boldsymbol H_{1}^{T}\quad
\boldsymbol H_{2}^{T}\quad \cdots \quad \boldsymbol H_{r}^{T}\quad
\cdots \quad \boldsymbol H_{M}^{T}]^{T}$, the total precoding matrix
can be obtained as $\boldsymbol P^{\rm ZF}=\boldsymbol
H^{H}(\boldsymbol H\boldsymbol H^{H})^{-1}$. The precoding matrix
$\boldsymbol P^{\rm ZF}$ can be expanded to $\boldsymbol P^{\rm
ZF}=[\boldsymbol P_{1}\quad \boldsymbol P_{2}\quad \cdots \quad
\boldsymbol P_{r}\quad \cdots \quad \boldsymbol P_{M}]$.
Simultaneously, the precoding matrix ${\boldsymbol P}_{r}'$ can be
generated from the null space of the $r$th user channel $\boldsymbol
H_{r}$ by singular value decomposition (SVD) \cite{mukherjee}. Note
that other precoding strategies
\cite{Gaojie,Xiaotao1,Keke1,Keke2,Keke3,sint} and MMSE filters can
also be considered
\cite{int,Chen,Meng,l1cg,zhaocheng,alt,jiolms,jiols,jiomimo,jidf,fa10,saabf,barc,honig,mswfccm,song,locsme}.
As a result, we have $\boldsymbol H_{r}{\boldsymbol
P}_{r}'=\boldsymbol 0 $, which means the jamming signal does not
interfere the user's signal. The received data for each user or
eavesdropper can be described by
\begin{equation}
{\boldsymbol y}_{r} =\beta_{r}^{-1}({\boldsymbol H}_{r}{\boldsymbol P}_{r}
{\boldsymbol s}_{r} +{\boldsymbol H}_{r}{\boldsymbol P}_{r}' {\boldsymbol
s}_{r}'+ {\boldsymbol H}_{r}\sum \limits_{j=1,j\neq r}^T {\boldsymbol P}_{j}
{\boldsymbol s}_{j}+ {\boldsymbol n}_{r}),\label{eqn:yr}
\end{equation}
where $\beta_{r}=\sqrt{\dfrac{E_{r}}{||\boldsymbol
P_{r}||+||\boldsymbol P_{r}'||}}$ is used to ensure that the
transmit power after precoding remains the same as the original
transmit power $E_{r}$ for user $r$.

\subsection{Secrecy Rate and Other Relevant Metrics}

In this subsection, we describe the main performance metrics used in the
literature to assess the performance of precoding algorithms.

\subsubsection{Secrecy rate and secrecy capacity}

According to \cite{wyner}, the level of secrecy is measured by the
uncertainty of Eve about the message $R_{e}$ which is called the
equivocation rate. With the total power equal to $E_{s}$, the
maximum secrecy capacity $C_{s}$ for the MIMO system without AN is
expressed as \cite{oggier}
\begin{equation}
\begin{split}
C_{s} &=\max_{{\boldsymbol Q}_{s}\geq 0,  \rm Tr({\boldsymbol Q}_{s}) =
E_{s}}\log(\det({\boldsymbol I}+{\boldsymbol H}_{ba} {\boldsymbol Q}_{s} {\boldsymbol H}_{ba}^H))\\
& \quad -\log(\det({\boldsymbol I}+ {\boldsymbol H}_{ea}{\boldsymbol Q}_{s} {\boldsymbol H}_{ea}^H)),
\end{split}
\label{eqn:Rs1}
\end{equation}
where the quantity ${\boldsymbol Q}_{s}$ is the covariance matrix
associated with the signal after precoding. However, the channels
between the transmitter and the eavesdroppers are usually not
perfectly known in reality. This situation is known as the imperfect
channel state information (CSI) case in \cite{goel}, which we will
address in our studies.

\subsubsection{Computational complexity}

According to the study in \cite{haardt}, non-linear precoding
techniques can approach the maximum channel capacity with high
computational complexity. High complexity of the algorithm directly
leads to a high cost of power consumption. In our research, however,
novel non-linear precoding algorithms with reduced complexity are
developed.

\subsubsection{BER performance}

Ideally, we would like the users to experience reliable communication and the
eavesdroppers to have a very high BER (virtually no reliability when
communicating). The algorithm is supposed to achieve high diversity for the MIMO
system.

\section{Review of the SO-THP Algorithm}
\label{sec:Review}

\begin{figure}[h]
\centering
\includegraphics[scale=0.43]{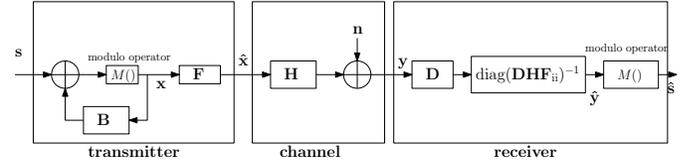}
\caption{Centralized SO-THP structure} \label{fig:dso}
\end{figure}

In this section, a brief review of the conventional successive
optimization THP (SO-THP) in \cite{stankovic} is given. The general
structure of the SO-THP algorithm is illustrated in Fig.
\ref{fig:dso} and its main implementation steps are introduced in
the following.

In Fig. \ref{fig:dso}, a modulo operation $M(\cdot)$ which is
defined in \cite{dietrich} is employed to fulfill the SO-THP
algorithm. Based on \cite{haardt}, THP can be equivalently
implemented in a successive block diagonalization manner. In
particular, the precoding matrix is given by
\begin{equation}
\boldsymbol P_{r}^{\rm BD}=\tilde{
\boldsymbol V_{r}}^{(0)} {\boldsymbol V_{\rm eff}}, \label{eqn:Pk}
\end{equation}
where $\tilde{ \boldsymbol V_{r}}^{(0)}\in {\mathbb{C}}^{N_{t}\times
N_{r}}$ is the nulling matrix of $r$th user's channel, ${\boldsymbol
V_{\rm eff}}$ is a unitary matrix of the corresponding effective
channel and the demodulation matrix of the $r$th user is chosen as
${\boldsymbol D_{r}}={\boldsymbol U_{\rm eff}^{H}}$, where
${\boldsymbol U_{\rm eff}^{H}}$ is also obtained from the effective
channel. Given a channel matrix $\tilde{\boldsymbol
H_{r}}=[\tilde{\boldsymbol H_{1}}^{T}\quad \tilde{\boldsymbol
H_{2}}^{T}\quad \cdots \quad \tilde{\boldsymbol H_{r-1}}^{T}\quad
\tilde{\boldsymbol H_{r+1}}^{T} \cdots \quad \tilde{\boldsymbol
H}_{T}^{T}]^{T}$, $\tilde{ \boldsymbol V_{r}}^{(0)}$ can be obtained
by the SVD operation $\tilde{\boldsymbol H_{r}}={\tilde{\boldsymbol
U}_{r}}{\tilde{\boldsymbol \Sigma}_{r}}[{\tilde{\boldsymbol
V}_{r}}^{(1)}{\tilde{\boldsymbol V}_{r}}^{(0)}]^H$. Based on
$\tilde{ \boldsymbol V_{r}}^{(0)}$, an effective channel can be
calculated and with second SVD operation ${\boldsymbol H}_{\rm
eff}={\boldsymbol H}_{r}\tilde{ \boldsymbol
V_{r}}^{(0)}={{\boldsymbol U}_{\rm eff}}{{\boldsymbol \Sigma}_{\rm
eff}}{{\boldsymbol V}_{\rm eff}}^H$ we are capable of getting
${\boldsymbol V_{\rm eff}}$ and ${\boldsymbol U_{\rm eff}^{H}}$. For
each iteration, the SO-THP algorithm selects the user with maximum
capacity from the remaining users and process it first. The
selection criterion is described as
\begin{equation}
\arg\min_{r}(C_{max,r}-C_{r}); \label{eqn:argk}
\end{equation}
where $C_{max,r}$ denotes the maximum capacity of the $r$th user and
$C_{r}$ is the capacity considering the interference from the other
users. If we assume there is no interference from other users and
the capacity can be achieved by the SVD procedure, we have
\begin{equation}
{\boldsymbol H}_{r}={{\boldsymbol U}_{r}}{{\boldsymbol
\Sigma}_{r}}[{{\boldsymbol V}_{r}}^{(1)}{{\boldsymbol
V}_{r}}^{(0)}]^H, \label{eqn:Hch1}
\end{equation}
\begin{equation}
C_{max,r}=\log_{2} {\det}\left(\boldsymbol I+{\boldsymbol
H}_{r}{{\boldsymbol V}_{r}}^{(1)}{{{{\boldsymbol
V}_{r}}^{(1)}}^H}{{{\boldsymbol H}_{r}}^H}\right). \label{eqn:Cmax}
\end{equation}
In the scenario considering the interference from the other users,
the BD decomposition is implemented on the channels of the remaining
users in each iteration:
 \begin{equation}
C_{r}=\log_{2} {\det}\left(\boldsymbol I+{\boldsymbol
H}_{r}{{\boldsymbol P}_{r}}{{{{\boldsymbol P}_{r}}}^H}{{{\boldsymbol
H}_{r}}^H}\right); \label{eqn:Ck}
\end{equation}
Therefore, the filters for the SO-THP algorithm can be obtained as
\begin{equation}
{\boldsymbol F}= \left( {\boldsymbol P}_{1}^{\rm BD}  \cdots  {\boldsymbol
P}_{T}^{\rm BD}  \right), \label{eqn:Fbd}
\end{equation}
\begin{equation}
{\boldsymbol D}=\begin{pmatrix}
 {\boldsymbol U}_{\rm eff1}^{H} &         & &  \\
         & \ddots & &  \\
         &         & {\boldsymbol U}_{\rm effT}^{H} &  \\
 \end{pmatrix},
\label{eqn:Dbd}
\end{equation}
\begin{equation}
{\boldsymbol B}={\rm lower~triangular}\left({\boldsymbol
D}{\boldsymbol H}{\boldsymbol F}\bullet {\rm
diag}\left([{\boldsymbol D}{\boldsymbol H}{\boldsymbol
F}]_{ii}^{-1}\right)\right). \label{eqn:Bbd}
\end{equation}
It is worth noting that ${\boldsymbol F}$ in (\ref{eqn:Fbd}),
${\boldsymbol D}$ in (\ref{eqn:Dbd}) are calculated in the reordered
way according to equation(5), and the scaling matrix $\boldsymbol
G={\rm diag}\left([{\boldsymbol D}{\boldsymbol H}{\boldsymbol
F}]_{ii}^{-1}\right)$.

\section{Proposed Precoding Algorithms}
\label{sec:Algorithms}

In this section, we present three non-linear precoding algorithms
SO-THP+GMI, SO-THP+S-GMI and LR-SO-THP+S-GMI for the downlink of
MU-MIMO systems, and a selection criterion based on capacity is
devised for these algorithms. We then derive filters for the three
proposed precoding techniques, which are computationally simpler
than SO-THP.

According to \cite{haardt}, the conventional SO-THP algorithm has
the advantage of improving the BER and the sum rate performances,
however, the complexity of this algorithm is high due to the
successive optimization procedure and the multiple SVD operations.
In \cite{hakjea}, an approach called generalized MMSE channel
inversion (GMI) is developed to overcome the noise enhancement
drawback of BD caused by its completely focus on the suppression of
multi-user interference. Later in \cite{keke1}, it has been shown
that the complete suppression of multi-user interference is not
necessary and residual interference is so small that cannot affect
the sum-rate performance. This approach is called simplified GMI
(S-GMI). The proposed algorithms are inspired by dirty paper coding
(DPC) \cite{costa} and other non-linear precoding techniques
\cite{haardt,huang,zuthp} which have been investigated for the
downlink of MU-MIMO systems.

\subsection{SO-THP+GMI Algorithm}

The proposed SO-THP+GMI algorithm mainly focuses on achieving high
secrecy rate performance with lower complexity than the conventional
SO-THP algorithm. In the conventional SO-THP algorithm the precoding
matrix as well as the receive filters are obtained with
(\ref{eqn:Pk}) using the BD algorithm without considering noise
enhancement. In \cite{hakjea}, the GMI scheme uses the QR
decomposition to decompose the MMSE channel inversion
$\bar{{\boldsymbol H}} \in {\mathbb{C}}^{N_{t}\times TN_{r}}$ as
expressed by
\begin{equation}
\bar{{\boldsymbol H}}=({{\boldsymbol H}^H}{\boldsymbol H}+\alpha \boldsymbol I)^{-1}{\boldsymbol H}^H,
\label{eqn:parh}
\end{equation}
\begin{equation}
\bar{{\boldsymbol H}_{r}}=[\bar{{\boldsymbol Q}_{r}}^{(0)} \quad \bar{{\boldsymbol Q}_{r}}^{(1)}]\bar{{\boldsymbol R}_{r}} \qquad  for \qquad r=1,\cdots,T, \label{eqn:parhqr}
\end{equation}
where $\bar {{\boldsymbol H}_{r}}\in {\mathbb{C}}^{N_{t}\times
N_{r}}$, $\bar{{\boldsymbol Q}_{r}}^{(0)} \in
{\mathbb{C}}^{N_{t}\times N_{t}}$ is an orthogonal matrix,
$\bar{{\boldsymbol R}_{r}}\in {\mathbb{C}}^{N_{t}\times N_{r}}$ is
an upper triangular matrix. In (\ref{eqn:parh}), the noise is taken
into account. As a result, the generation of the precoding matrix
will mitigate the noise enhancement. When the GMI generated
precoding matrix is used to calculate the channel capacity with
(\ref{eqn:Ck}), the reduced noise contributes to the increase of
secrecy rate. Also with (\ref{eqn:parhqr}), the QR decomposition
reduces the computational complexity as compared with the
conventional SO-THP algorithm implementing the SVD decomposition. To
completely mitigate the interference,a transmit combining matrix
${\boldsymbol T}_{r}$ given in \cite{hakjea} is applied to
$\bar{{\boldsymbol Q}_{r}}^{(0)}$. Once we have $\bar{{\boldsymbol
Q}_{r}}^{(0)}$ and ${\boldsymbol T}_{r}$, we can write the relation
\begin{equation}
{\boldsymbol H}_{r}\bar{\boldsymbol Q}_{r}^{(0)}{\boldsymbol
T}_{r}={\bar{\boldsymbol U}}_{r}{\bar{{\boldsymbol
\Sigma}_{r}}}{\bar{{\boldsymbol V}_{r}}}^H, \label{eqn:Trsvd}
\end{equation}
Then the precoding matrix as well as the receive filter for the GMI
scheme are given by
\begin{equation}
{\boldsymbol P}_{\rm GMI} = [\bar{\boldsymbol
Q}_{1}^{(0)}{\boldsymbol T}_{1}\bar{\boldsymbol V}_{1} \quad
\bar{\boldsymbol Q}_{2}^{(0)}{\boldsymbol T}_{2}\bar{\boldsymbol
V}_{2} \quad \cdots \quad \bar{\boldsymbol Q}_{T}^{(0)} {\boldsymbol
T}_{T} \bar{\boldsymbol V}_{T}], \label{eqn:pgmi}
\end{equation}
\begin{equation}
{\boldsymbol M}_{\rm GMI} = {\rm diag}\{ \bar{\boldsymbol U}_{1}^H
\quad \bar{\boldsymbol U}_{2}^H \quad \cdots \quad \bar{\boldsymbol
U}_{T}^H \}, \label{eqn:mgmi}
\end{equation}
where ${\boldsymbol P}_{\rm GMI}\in {\mathbb C}^{N_{t} \times
N_{t}}$ and  ${\boldsymbol M}_{\rm GMI} \in {\mathbb C}^{N_{t}
\times N_{t}}$. The details of the proposed SO-THP+GMI algorithm to
obtain the precoding and receive filter matrices are given in the
table of Algorithm 1.

\begin{algorithm}
\caption{Proposed SO-THP+GMI Precoding}
\begin{algorithmic}[1]
\FOR {$r=1:T$} \STATE ${\boldsymbol G}_{r}={\boldsymbol H}_{r};$
\STATE ${\boldsymbol G}_{r}={{\boldsymbol U}_{r}}{ {\boldsymbol
\Sigma}_{r}}[{{\boldsymbol V}_{r}}^{(1)}{{\boldsymbol
V}_{r}}^{(0)}]^H;$ \STATE ${\boldsymbol F}_{r}={{\boldsymbol
V}_{r}}^{(1)};$ \STATE{ $C_{max,r}=\log_{2}\det\left(\boldsymbol
I+{\boldsymbol R}_{k,r}^{-1} {\boldsymbol G}_{r}{\boldsymbol
F}_{r}{{{\boldsymbol F}_{r}}^H}{{{\boldsymbol G}_{r}}^H}\right);$ }
\ENDFOR \STATE ${\boldsymbol M}={\boldsymbol H};$ \LOOP \WHILE
{$r=T:1$} \FOR {$n=1:r$} \STATE {${\boldsymbol G}=({{\boldsymbol
M}^H}{\boldsymbol M}+\alpha \boldsymbol I)^{-1}{\boldsymbol M}^H$}
\STATE ${\boldsymbol G}_{n}=[\bar{{\boldsymbol Q}_{r}}^{(0)} \quad
\bar{{\boldsymbol Q}_{r}}^{(1)}]\bar{{\boldsymbol R}_{n}}$
\STATE
${{\boldsymbol M}_{n}}\bar{{\boldsymbol Q}_{r}}^{(0)}{\boldsymbol
T_{r}}={\tilde{{\boldsymbol U}_{n}}}{\tilde{{\boldsymbol
\Sigma}_{n}}}{\tilde{{\boldsymbol V}_{n}}}^H$
\STATE ${{\boldsymbol
P}_{n}}=\bar{{\boldsymbol Q}_{n}}^{(0)}{\boldsymbol
T_{r}}{\tilde{{\boldsymbol V}_{n}}^{(1)}}$
\ENDFOR
\FOR {$j=1:r$}
\STATE {$C_{j}=\log_{2} \det\left(\boldsymbol I+{\boldsymbol
R}_{k,j}^{-1} {\boldsymbol M}_{j}{\boldsymbol P}_{j}{{{\boldsymbol
P}_{j}}^H}{{{\boldsymbol M}_{j}}^H}\right);$}
\ENDFOR
\STATE
{$a_{r}=\arg\min_{j}(C_{max,j}-C_{j});$}
\STATE ${\boldsymbol
F}_{r}={\boldsymbol P}_{a_{r}};$
\STATE ${\boldsymbol
D}_{r}=\tilde{{\boldsymbol U}_{a_{r}}}^{H};$
\STATE ${\boldsymbol
M}=[{{\boldsymbol H}_{1}}^T \cdots {{\boldsymbol H}_{a_{r}-1}}^T
{{\boldsymbol H}_{a_{r}+1}}^T \cdots {{\boldsymbol H}_{T}}^T]^T$
\ENDWHILE \ENDLOOP
\STATE ${\boldsymbol F}= \left( {\boldsymbol
F}_{1}  \cdots  {\boldsymbol F}_{T}  \right);$
\STATE ${\boldsymbol
D}=\begin{pmatrix}
 {\boldsymbol D}_{1} &         & &  \\
         & \ddots & &  \\
         &         & {\boldsymbol D}_{T} &  \\
 \end{pmatrix} $
\STATE ${\boldsymbol B}={\rm lower~triangular}\left({\boldsymbol D}{\boldsymbol H}{\boldsymbol F}\bullet {\rm diag}\left([{\boldsymbol D}{\boldsymbol H}{\boldsymbol F}]_{rr}^{-1}\right)\right)$
\end{algorithmic}
\end{algorithm}

\subsection{SO-THP+S-GMI Algorithm}

Further development on SO-THP+GMI with complexity reduction leads to
a novel SO-THP+S-GMI alogrithm. A simplified GMI (S-GMI) has been
developed in \cite{keke1} as an improvement of the original RBD
precoding in \cite{stankovic}. This is known as S-GMI. In
(\ref{eqn:Trsvd}), a transmit combining matrix is applied to achieve
complete interference cancelation between different users. In this
case, the interference will not be completely mitigated, resulting
in a slight decrease of the sum-rate even though the complexity will
have a significant reduction \cite{keke1}. Here, we incorporate the
S-GMI technique into an SO-THP scheme and devise the SO-THP+S-GMI
algorithm. The transmit and receive filters of the proposed
SO-THP+S-GMI algorithm are described by
\begin{equation}
{{\boldsymbol H}_{r}}\bar{{\boldsymbol Q}_{r}}^{(0)} = \tilde{{\boldsymbol
U}_{r}}\tilde{{\boldsymbol \Sigma}_{r}}\tilde{{{\boldsymbol
V}_{r}}}^H, \label{eqn:sgmisvd}
\end{equation}
\begin{equation}
{{\boldsymbol P}_{\rm S-GMI}}=[\bar{{\boldsymbol Q}_{1}}^{(0)}
\tilde{{\boldsymbol V}_{1}} \quad \bar{{\boldsymbol
Q}_{2}}^{(0)}\tilde{{\boldsymbol V}_{2}}\quad \cdots \quad
\bar{{\boldsymbol Q}_{T}}^{(0)}\tilde{{\boldsymbol V}_{T}}],
\label{eqn:psgmi}
\end{equation}
\begin{equation}
{{\boldsymbol M}_{\rm S-GMI}}={\rm diag} \{\tilde{{\boldsymbol U}_{1}}^H
\quad \tilde{{\boldsymbol U}_{2}}^H \quad \cdots \quad
\tilde{{\boldsymbol U}_{T}}^H\}, \label{eqn:msgmi}
\end{equation}
where ${{\boldsymbol P}_{\rm S-GMI}}\in {\mathbb{C}}^{N_{t}\times N_{t}}$,
${{\boldsymbol M}_{\rm S-GMI}}\in {\mathbb{C}}^{N_{t}\times N_{t}}$.

With reduced computational complexity, the SO-THP+S-GMI algorithm is capable of
achieving better secrecy rate performance especially at lower SNR. The detailed S-GMI
procedure implemented in the proposed SO-THP+S-GMI algorithm is shown in
Algorithm 2. Cooperated with Algorithm 1, the precoding and receive filter matrices can be obtained.

\begin{algorithm}
\caption{S-GMI Precoding}
\centering
\begin{algorithmic}[1]
\FOR {$n=1:r$}
\STATE ${\boldsymbol G}=({{\boldsymbol M}^H}{\boldsymbol M}+\alpha \boldsymbol I)^{-1}{\boldsymbol M}^H$
\STATE ${\boldsymbol G}_{n}=[\bar{{\boldsymbol Q}_{n}}^{(0)} \quad \bar{{\boldsymbol Q}_{n}}^{(1)}]\bar{{\boldsymbol R}_{n}}$
\STATE ${{\boldsymbol M}_{n}}\bar{{\boldsymbol Q}_{n}}^{(0)}={\tilde{{\boldsymbol U}_{n}}}{\tilde{{\boldsymbol \Sigma}_{n}}}{\tilde{{\boldsymbol V}_{n}}}^H$
\STATE ${{\boldsymbol P}_{n}}=\bar{{\boldsymbol Q}_{n}}^{(0)}{\tilde{{\boldsymbol V}_{n}}^{(1)}}$
\ENDFOR
\end{algorithmic}
\end{algorithm}

\subsection{LR-SO-THP+S-GMI Algorithm}

The development in linear algebra contribute to the lattice
reduction technique application in wireless networks. According to
study in \cite{yao}, a basis change may lead to improved performance
as corroborated by lattice reduction techniques. The more correlated
the columns of channel $\boldsymbol H$, the more significant the
improvements will be. To achieve full diversity of the system, with
complex lattice reduction algorithm (CLR) \cite{keke2}, the LR
transformed channel for the $r$th user is obtained as
\begin{equation}
{{\boldsymbol H}_{red_{r}}^{H}}={{\boldsymbol H}_{r}^{H}}{{\boldsymbol L}_{r}}
\label{eqn:Hredr}
\end{equation}
where ${{\boldsymbol H}_{red_{r}}}\in {\mathbb{C}}^{N_{r}\times
N_{t}}$ is the transposed reduced channel matrix. The quantity
${{\boldsymbol L}_{r}} \in {\mathbb{C}}^{N_{r}\times N_{r}}$ is the
transform matrix generated by the CLR algorithm. Note that the
transmit power constraint is satisfied since ${\boldsymbol M}_{r}$
is a unimodular matrix.

Compared to the conventional SO-THP algorithm, the lattice reduced
channel matrix $\boldsymbol H_{red_{n}}$ is employed in the
conventional S-GMI algorithm. The details of the LR aided S-GMI Procedure are
given in Algorithm 3. Cooperated with Algorithm 1, we can complete the calculation of precoding and receive filter matrices.

\begin{algorithm}
\caption{Lattice-Reduction aided S-GMI Procedure}
\centering
\begin{algorithmic}[1]
\FOR {$n=1:r$}
\STATE ${\boldsymbol G}=({{\boldsymbol H}^H}{{\boldsymbol H}}+\alpha \boldsymbol I)^{-1}{{\boldsymbol H}}^H$
\STATE $[{\boldsymbol H_{{red}_{n}}}^{H} \quad {\bar{\boldsymbol Q}_{n}}^{(0)}]=\rm CLLL({\boldsymbol G}_{n}^{H})$
\STATE ${{\boldsymbol M}_{n}}=\boldsymbol H_{{red}_{n}}$
\STATE ${{\boldsymbol M}_{n}}\bar{{\boldsymbol Q}_{n}}^{(0)}={\tilde{{\boldsymbol U}_{n}}}{\tilde{{\boldsymbol \Sigma}_{n}}}{\tilde{{\boldsymbol V}_{n}}}^H$
\STATE ${{\boldsymbol P}_{n}}=\bar{{\boldsymbol Q}_{n}}^{(0)}{\tilde{{\boldsymbol V}_{n}}^{(1)}}$
\ENDFOR
\end{algorithmic}
\end{algorithm}

\section{Analysis of the Algorithms}
\label{sec:Analysis}

In this section, we develop an analysis of the secrecy rate of the proposed
precoding algorithms along with a comparison of the computational complexity
between the proposed and existing techniques.

\subsection{Computational Complexity Analysis}
\begin{table}[ht]
\caption{Computational complexity of the proposed SO-THP+GMI algorithm } 
\centering
\setlength{\tabcolsep}{0pt}
\begin{tabular}{c| c| c| c}
\hline\hline
Steps & Operations & Flops & Case \\ [0.5ex]
 &  &  & {$(2,2,2)\times 6$} \\ [0.5ex]
\hline
1&${\boldsymbol G}_{r}={{\boldsymbol U}_{r}}{{\boldsymbol \Sigma}_{r}}[{{\boldsymbol V}_{r}}^{(1)}{{\boldsymbol V}_{r}}^{(0)}]^H;$&$32R(N_{t}N_{r}^2$&\\
&&$+N_{r}^3)$ & 3072 \\
2 & $\bar {\boldsymbol G}=$ & $(2N_{t}^3-2N_{t}^2$&\\
&${\boldsymbol G}=({{\boldsymbol H}^H}{\boldsymbol H}+\alpha \boldsymbol I)^{-1}{\boldsymbol H}^H$&$+N_{t}+16N_{R} N_{t}^2)$ & 3822 \\
3 & $\bar {\boldsymbol G}_{n}=\bar{{\boldsymbol Q}_{n}}\bar{{\boldsymbol R}_{n}}$ & $\sum\limits_{r=1}^R 16r(N_{t}^2 N_{r}$&\\
&&$ + N_{t} N_{r}^2 +\frac{1}{3} N_{r}^3  )$ & 9472 \\
4 & ${{\boldsymbol H}_{eff,n}}={{\boldsymbol H}_{n}}\bar{{\boldsymbol Q}_{n}}{\boldsymbol T}_{n}$ & $\sum\limits_{r=1}^R 16r N_{R} N_{t}^{2}$ & 20736 \\
5 & ${{\boldsymbol H}_{eff,n}}={{\boldsymbol U}_{n}^{(4)}}{{\boldsymbol \Sigma}_{n}^{(4)}}{{{\boldsymbol V}_{n}}^{(4)}}^H$ & $\sum\limits_{r=1}^R 64r(\frac{9}{8}N_{r}^{3}+ $&\\
&&$N_{t} N_{r}^{2}+\frac{1}{2}N_{t}^{2} N_{r})$ & 26496 \\ [1ex]
6 & ${\boldsymbol B}={\rm lower~triangular}$&&\\
& $\left({\boldsymbol D}{\boldsymbol H}{\boldsymbol F}\bullet {\rm diag}\left([{\boldsymbol D}{\boldsymbol H}{\boldsymbol F}]_{rr}^{-1}\right)\right)$ & $ 16N_{R} N_{t}^{2}$ & 3456 \\ [1ex]
\hline
&&&total 67054\\
\hline\hline
\end{tabular}
\end{table}

According to \cite{keke2}, it can be calculated that the cost of the QR in
FLOPs is $22.4\%$ lower than BD. The results shown in Table I indicate that the
complexity is reduced by about $22.4\%$ by the proposed SO-THP+GMI compared
with the conventional SO-THP calculated in the same way. Based on the proposed
SO-THP+GMI algorithm, further complexity reduction can be achieved by
SO-THP+S-GMI and the complexity is about $34.4\%$ less than that of the
conventional SO-THP algorithm.

\begin{center}
\begin{figure}[h]
\centering
\includegraphics[width=\linewidth]{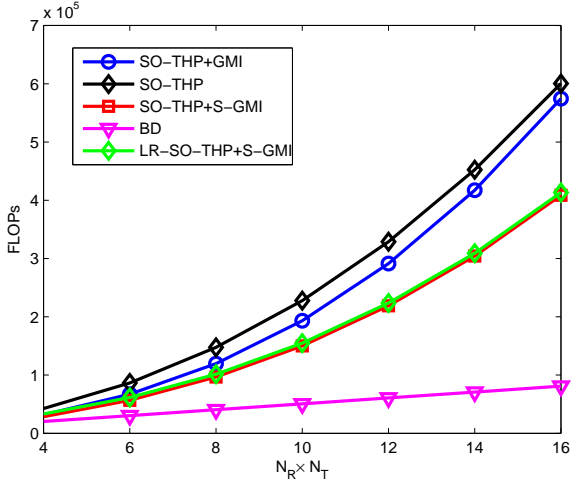}
\caption{Computational complexity in FLOPs for MU-MIMO systems}
\label{fig:com}
\end{figure}
\end{center}

Fig. \ref{fig:com} shows the required FLOPS of the proposed and
existing precoding algorithms. Linear precoding gives lower
computational complexity but the BER performance is worse than
non-linear ones. The three proposed algorithms show an advantage
over the conventional SO-THP algorithm in terms of complexity. Among
all the three proposed algorithms, SO-THP+S-GMI has the lowest
complexity followed by LR-SO-THP+S-GMI. SO-THP+GMI requires the
highest complexity.

\subsection{Secrecy rate Analysis}

 \emph{Theorem} 1:
In full-rank MU-MIMO systems with perfect knowledge of CSI, the proposed
algorithms are capable of achieving a high secrecy rate and in the high-SNR
regime (i.e., $E_{s} \rightarrow \infty$) the secrecy rate will converge to
$C_{sec}^{E_{s}\rightarrow \infty}$ which is given as (\ref{eqn:Csec}),
\begin{equation}
C_{sec}^{E_{s}\rightarrow \infty}=\log\Bigg(\det \Big(({{\boldsymbol H}_{ea}{\boldsymbol H}_{ea}^H
})^{-1}({{\boldsymbol H}_{ba}{\boldsymbol H}_{ba}^H })\Big)\Bigg)
\label{eqn:Csec}
\end{equation}
\begin{proof}
Under the conditions
\begin{equation}
{\boldsymbol H}_{ba}^H {\boldsymbol
H}_{ba} \succeq {\boldsymbol H}_{ea}^H {\boldsymbol H}_{ea}
\end{equation}
\begin{equation}
\rm rank({\boldsymbol H}_{ba})=\rm rank({\boldsymbol H}_{ea})
\end{equation}
and based on (\ref{eqn:Rs1}) we can have the secrecy capacity
expressed as (\ref{eqn:lag51}). If $\Gamma({\boldsymbol
P})={({\boldsymbol H}_{ea}\boldsymbol P \boldsymbol P^H{\boldsymbol
H}_{ea}^H )^{-1}}{({\boldsymbol H}_{ba}\boldsymbol P \boldsymbol
P^H{\boldsymbol H}_{ba}^H )}$, (\ref{eqn:lag51}) can be converted to
(\ref{eqn:lag5}).
\begin{table*}
\centering
\begin{minipage}{\textwidth}
\hrule
\begin{equation}
C_{s}=\max_{{\boldsymbol Q}_{s}\geq 0, \rm Tr({\boldsymbol Q}_{s}) =
E_{s}}\log\Bigg(\det\Big({({\boldsymbol I}+ {\boldsymbol
H}_{ea}{\boldsymbol Q}_{s} {\boldsymbol
H}_{ea}^H)^{-1}}{({\boldsymbol I}+{\boldsymbol H}_{ba} {\boldsymbol
Q}_{s} {\boldsymbol H}_{ba}^H)}\Big)\Bigg) \label{eqn:lag51}
\end{equation}
\begin{equation}
C_{s}= \max_{{\boldsymbol Q}_{s}\geq 0, \rm Tr({\boldsymbol Q}_{s})= E_{s}}\log\Bigg(\det\Big(\Gamma_{ba}({\boldsymbol P}) \Phi_{1}({\boldsymbol P},{\boldsymbol Q_{s}})^{-1}\Phi_{2}({\boldsymbol P},{\boldsymbol Q_{s}})\Big)\Bigg)
\label{eqn:lag5}
\end{equation}
\begin{equation}
\Phi_{1}({\boldsymbol P},{\boldsymbol Q_{s}})={({\boldsymbol H}_{ea}\boldsymbol P \boldsymbol P^H{\boldsymbol H}_{ea}^H )^{-1}}{({\boldsymbol I}+ {\boldsymbol H}_{ea}{\boldsymbol Q}_{s} {\boldsymbol H}_{ea}^H)}
\label{eqn:lag52}
\end{equation}
\begin{equation}
\Phi_{2}({\boldsymbol P},{\boldsymbol Q_{s}})={{({\boldsymbol H}_{ba}\boldsymbol P \boldsymbol P^H{\boldsymbol H}_{ba}^H)^{-1}}{({\boldsymbol I}+{\boldsymbol H}_{ba} {\boldsymbol Q}_{s} {\boldsymbol H}_{ba}^H)}}
\label{eqn:lag53}
\end{equation}
\hrule
\end{minipage}
\end{table*}

In (\ref{eqn:lag5}), (\ref{eqn:lag52}) and (\ref{eqn:lag53}), $\boldsymbol P$ is the precoding matrix derived from the legitimate users' channel.
With $\boldsymbol Q_{s}=E[\boldsymbol
x_{s} \boldsymbol x_{s}^{H}]=E[\boldsymbol P \boldsymbol s
\boldsymbol s^{H} \boldsymbol P^{H}]$, $E[\boldsymbol
s \boldsymbol s^{H}]=E_{s}$ and $\boldsymbol
P \boldsymbol P^{H}=\boldsymbol I$, we can have,
\begin{equation}
E[{(\boldsymbol P \boldsymbol P^{H})^{-1}}{\boldsymbol Q_{s}}]=E_{s}
\label{eqn:lag6}
\end{equation}
Then
\begin{equation}
E[{({\boldsymbol H}_{ba}\boldsymbol P \boldsymbol P^{H}{\boldsymbol H}_{ba}^H)^{-1}}{{\boldsymbol H}_{ba}\boldsymbol Q_{s}{\boldsymbol H}_{ba}^H}]=E_{s}
\label{eqn:lag7}
\end{equation}
\begin{equation}
E[{({\boldsymbol H}_{ea}\boldsymbol P \boldsymbol P^{H}{\boldsymbol H}_{ea}^H)^{-1}}{{\boldsymbol H}_{ea}\boldsymbol Q_{s}{\boldsymbol H}_{ea}^H}]=E_{s}
\label{eqn:lag71}
\end{equation}
In (\ref{eqn:lag5}), the expectation value is given as (\ref{eqn:lag8}).
Substituting (\ref{eqn:lag7}) into (\ref{eqn:lag8}) the formula can be
expressed as (\ref{eqn:lag9}).
\begin{table*}
\centering
\begin{minipage}{\textwidth}
\hrule
\begin{equation}
\begin{split}
SA&=E\Big[\Phi_{1}({\boldsymbol P},{\boldsymbol Q_{s}})^{-1}\Phi_{2}({\boldsymbol P},{\boldsymbol Q_{s}})\Big]\\
&=E\bigg[\Phi_{3}({\boldsymbol P},{\boldsymbol
Q_{s}})^{-1}\Big({({{\boldsymbol H}_{ba}\boldsymbol P  \boldsymbol
P^H{\boldsymbol H}_{ba}^H})^{-1}+({{\boldsymbol H}_{ba}\boldsymbol P
\boldsymbol P^H{\boldsymbol H}_{ba}^H})^{-1}{{\boldsymbol
H}_{ba}\boldsymbol Q_{s}{\boldsymbol H}_{ba}^H}})\Big)\bigg]
\label{eqn:lag8}
\end{split}
\end{equation}

\begin{equation}
\Phi_{3}({\boldsymbol P},{\boldsymbol Q_{s}})={({{\boldsymbol
H}_{ea}\boldsymbol P \boldsymbol P^H{\boldsymbol H}_{ea}^H
})^{-1}+({{\boldsymbol H}_{ea}\boldsymbol P \boldsymbol
P^H{\boldsymbol H}_{ea}^H })^{-1}{{\boldsymbol H}_{ea}\boldsymbol
Q_{s}{\boldsymbol H}_{ea}^H}} \label{eqn:lag81}
\end{equation}

\begin{equation}
\begin{split}
SA&=E\bigg[\Big({({{\boldsymbol H}_{ea}\boldsymbol P \boldsymbol P^H{\boldsymbol H}_{ea}^H })^{-1}+E_{s}}\Big)^{-1}\Big({({{\boldsymbol H}_{ba}\boldsymbol P \boldsymbol P^H{\boldsymbol H}_{ba}^H})^{-1}+E_{s}}\Big)\bigg]\\
&=E\bigg[\Big({({{\boldsymbol H}_{ea}\boldsymbol P \boldsymbol P^H{\boldsymbol H}_{ea}^H})^{-1}+E_{s}}+{\boldsymbol I}\Big)^{-1}\Big({({{\boldsymbol H}_{ba}\boldsymbol P \boldsymbol P^H{\boldsymbol H}_{ba}^H})^{-1}-({{\boldsymbol H}_{ea}\boldsymbol P \boldsymbol P^H {\boldsymbol H}_{ea}^H })^{-1}}\Big)\bigg]
\label{eqn:lag9}
\end{split}
\end{equation}
\hrule
\end{minipage}
\end{table*}

According to (\ref{eqn:lag9}), in the high-SNR regime and when $\rm
SNR\rightarrow \infty$, $E_{s}\rightarrow \infty$, $SA \rightarrow \boldsymbol
I$. Then, the secrecy rate expressed in (\ref{eqn:lag5}) will result in (\ref{eqn:lag10}).
\begin{table*}
\centering
\begin{minipage}{\textwidth}
\hrule
\begin{equation}
C_{sec}^{E_{s}\rightarrow \infty}= \log\Bigg(\det \Big(({{\boldsymbol H}_{ea}\boldsymbol P \boldsymbol P^H{\boldsymbol H}_{ea}^H })^{-1}({{\boldsymbol H}_{ba}\boldsymbol P \boldsymbol P^H{\boldsymbol H}_{ba}^H })\Big)\Bigg)
\label{eqn:lag10}
\end{equation}
\hrule
\end{minipage}
\end{table*}
To satisfy the power constrain, we always have $E[\boldsymbol P \boldsymbol
P^H]=\boldsymbol I$, then the secrecy rate $C_{sec}$ will converge to a
constant, that is,
\begin{equation}
C_{sec}^{E_{s}\rightarrow \infty}= \log \bigg(\det({{\boldsymbol H}_{ea}{\boldsymbol H}_{ea}^H })^{-1}({{\boldsymbol H}_{ba}{\boldsymbol H}_{ba}^H
}) \bigg)
\label{eqn:cs1}
\end{equation}
This completes the proof.
\end{proof}
In the following, the percentage of the injected artificial noise
power is set to 40\% of the total transmit power. When AN is added
during the transmission, equation (\ref{eqn:Rs1}) can be transformed
to:
\begin{equation}
\begin{split}
&\log\Big(\det({\boldsymbol I}+{\boldsymbol H}_{ba} {\boldsymbol Q}_{s} {\boldsymbol H}_{ba}^H)\Big)\\
&-\log\bigg(\det\Big({\boldsymbol I}+ (\boldsymbol I+{\boldsymbol H}_{ea}{\boldsymbol Q}_{s}' {\boldsymbol H}_{ea}^H)^{-1}({\boldsymbol H}_{ea}{\boldsymbol Q}_{s} {\boldsymbol H}_{ea}^H)\Big)\bigg)
\label{eqn:Csan1}
\end{split}
\end{equation}
To assess the influence of different channel gain ratios between legitimate
users and the eavesdroppers, we fix the legitimate users' channel gain and
change the eavesdroppers'. The above equation (\ref{eqn:Csan1}) can be further
transformed to
\begin{equation}
\begin{split}
&\log(\det({\boldsymbol I}+{\boldsymbol H}_{ba} {\boldsymbol Q}_{s} {\boldsymbol H}_{ba}^H))\\
&-\log(\det({\boldsymbol I}+ (({{\boldsymbol H}_{ea}{\boldsymbol H}_{ea}^H})^{-1}+{\boldsymbol Q}_{s}')^{-1}{\boldsymbol Q}_{s})
\label{eqn:Csan2}
\end{split}
\end{equation}
In the high-SNR regime, $E_{s}\rightarrow\infty$, according to
(\ref{eqn:lag6}), $Q_{s}, Q_{s}' \rightarrow\infty$, the term $({{\boldsymbol
H}_{ea}{\boldsymbol H}_{ea}^H})^{-1}$ then can be omitted and the result is the
following expression
\begin{equation}
\log(\det({\boldsymbol I}+{\boldsymbol H}_{ba} {\boldsymbol Q}_{s} {\boldsymbol H}_{ba}^H))\\
-\log(\det({\boldsymbol I}+ ({\boldsymbol Q}_{s}')^{-1}{\boldsymbol Q}_{s}),
\label{eqn:Csan3}
\end{equation}
Considering artificial noise, $({\boldsymbol Q}_{s}')^{-1}{\boldsymbol
Q}_{s}=\rho/(1-\rho)\boldsymbol I$. When $\rho$ is fixed, then
$\log(\det({\boldsymbol I}+ ({\boldsymbol Q}_{s}')^{-1}{\boldsymbol Q}_{s}))$
would be a constant. From (\ref{eqn:Csan3}), the secrecy rate will increase
even when the eavesdroppers have better statistical channel knowledge than the
legitimate users.

\section{Simulation Results}
\label{sec:Simulation}

A system with $N_{t}=4$ transmit antennas and $T=2$ users as well as $K=1,2$
eavesdroppers is considered. Each user or eavesdropper is equipped with
$N_{r}=2$ and $N_{k}=2$ receive antennas.
$m=\frac{\sigma_{ea}^{2}}{\sigma_{ba}^{2}}$ represents the gain ratio between
the main and wire-tap channel.

\begin{figure}[h]
\centering
\includegraphics[width=\linewidth]{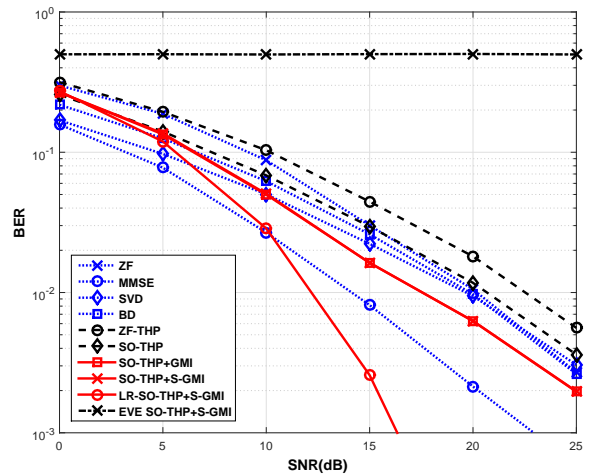}
\caption{BER performance with precoding techniques in $4 \times 4
\times 2$ MU-MIMO broadcast channel, $m=0.5$} \label{fig:ber1}
\end{figure}

\subsection{Perfect Channel State Information}


\begin{figure}[ht]
\centering
{\includegraphics[width=.45\linewidth]{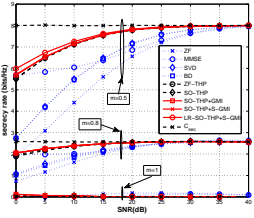}}\caption{Secrecy
rate performance with precoding techniques in $4 \times 4 \times 4$
MU-MIMO broadcast channel}\label{Sec}
\end{figure}

\begin{figure}[ht]
\centering \caption{Secrecy rate change with different artificial
noise power ratio}
{\includegraphics[width=.45\linewidth]{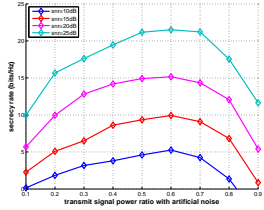}}\label{Sec}
\end{figure}

In Fig. \ref{fig:ber1} the proposed LR-SO-THP+S-GMI algorithm has the best BER
performance. In Fig. \ref{fig:442P}, in the scenario where $T>K$ the secrecy
rate of the proposed algorithms have around 5 bits/Hz higher rate than the
other precoding techniques. When $T=K$, Fig. \ref {Sec}(a) shows that the
proposed algorithms achieve a higher secrecy rate than the other techniques at
low SNRs. And the secrecy rate will converge to a constant which will depend on
the gain ratio between the main and the wire-tap channels $m$.

\subsection{Imperfect Channel State Information}

In the simulations, the channel errors are modeled as a complex random Gaussian
noise matrix $\boldsymbol E$ following the distribution
$\mathcal{CN}(0,\sigma_{e}^{2})$. Then, the imperfect channel matrix
$\boldsymbol H^{e}$ is defined as
\begin{equation}
\boldsymbol H^{e}=\boldsymbol H+\boldsymbol E
\label{eqn:Hbae}
\end{equation}
We assume the channels of the legitimate users are perfect and the eavesdropper
will have imperfect CSI.

\begin{figure}[ht]
\centering
{\includegraphics[width=.45\linewidth]{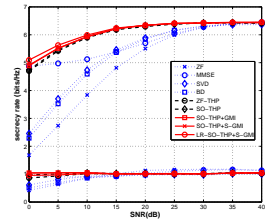}}\caption{Secrecy
rate with precoding techniques $4 \times 4 \times 4$ MU-MIMO
broadcast channel with imperfect CSI}\label{Sec}
\end{figure}

\begin{figure}[ht]
\centering
{\includegraphics[width=.45\linewidth]{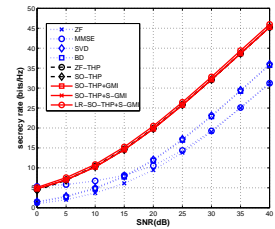}}\caption{Secrecy
rate with precoding techniques $4 \times 4 \times 2$ MU-MIMO
broadcast channel with imperfect CSI, AN and $m=2$}\label{Sec}
\end{figure}

In Fig. \ref {Sec}(b), the secrecy rate performance is evaluated in
the imperfect CSI scenario
\cite{landau2014robust,peng2016adaptive,ruan2014robust}. Compared
with the secrecy rate performance in Fig. \ref {Sec}(a), the secrecy
rate will suffer a huge decrease in the imperfect CSI scenario. When
$T=K$, Fig. \ref {Sec}(b) shows that the secrecy rate at low SNR is
degraded and the secrecy rate requires very high SNR to converge to
a constant. It is worth noting that the proposed SO-THP+S-GMI has
the best secrecy rate performance amongst the studied precoding
techniques.

\subsection{Imperfect Channel State Information With Artificial Noise}

In Fig. \ref{Sec}(c) AN is added and the total transmit power $E_{s}$ is the
same as before. Fig. \ref{Sec}(d) shows the secrecy rate with the change of the
transmit signal power ratio to the artificial noise. According to the secrecy
performance of Fig. \ref{Sec}(d), $40\%$ of the transmit power $E_{s}$ is used
to generate AN.

\section{Conclusion}
\label{sec:Conclusion}

Precoding techniques are widely used in the downlink of MU-MIMO wireless
networks to achieve good BER performance. They also contribute to the
improvement of the secrecy rate in physical layer. Among all the studied
non-linear precoding techniques, the proposed SO-THP+S-GMI algorithm requires
the lowest computational complexity which results in a significant improvement
on the efficiency. The BER and the secrecy rate performances of the
SO-THP+S-GMI algorithm are also superior to the existing linear and non-linear
algorithms considered.

\bibliographystyle{IEEEtran}
\bibliography{reference_eurasip}

\end{document}